\documentstyle[12pt]{article}
\begin{document}
\newcommand{\be}{\begin{equation}}
\newcommand{\ee}{\end{equation}}
\newcommand{\ba}{\begin{array}}
\newcommand{\ea}{\end{array}}
\newcommand{\bea}{\begin{eqnarray}}
\newcommand{\eea}{\end{eqnarray}}

\def\onehalf{{\textstyle \frac12}}
\def\tsty#1#2{{\textstyle\frac{#1}{#2}}}
\def\trans{^{\scriptscriptstyle\top}}
\def\ie{{\it i.e.}}
\def\Sp#1{{\rm Sp}(#1,\Re)}
\def\sp#1{{\rm sp}(#1,\Re)}
\def\SO#1{{\rm SO}(#1)}
\def\so#1{{\rm so}(#1)}
\def\SU#1{{\rm SU}(#1)}
\def\su#1{{\rm su}(#1)}
\def\uni#1#2{\hbox{#1($#2$)}}
\def\ket#1{\,\vert{#1}\rangle}
\def\bra#1{\langle{#1}\vert}
\def\xx{{\bf x}}\def\yy{{\bf y}}\def\zz{{\bf z}}
\def\pp{{\bf p}}\def\nn{{\bf n}}
\def\braket#1#2{\langle{#1}\vert{#2}\rangle}
\def\pmbf#1{\rlap{$#1$}{}\hskip-0.5pt{#1}}
\def\abs#1{{\scriptstyle|}#1{\scriptstyle|}}
\def\aabs#1{{\scriptscriptstyle|}#1{\scriptscriptstyle|}}
\def\tsty#1#2{{\textstyle\frac{#1}{#2}}}
\def\of#1{{{\scriptstyle(}#1{\scriptstyle)}}}
\def\oof#1{{{\scriptscriptstyle(}#1{\scriptscriptstyle)}}}
\def\brof#1{{{\scriptstyle[}#1{\scriptstyle]}}}
\def\totder#1#2{\frac{d #1}{d #2}}
\def\parder#1#2{\frac{\partial #1}{\partial #2}}
\def\Poi#1#2{\{#1,#2\}}
\def\ssr#1{{\scriptscriptstyle\rm #1}}
\def\ssty#1{{\scriptscriptstyle #1}}
\def\vecdos#1#2{\left({\matrix{#1\cr #2\cr}}\right)}
\def\matdos#1#2#3#4{\left({\matrix{#1&#2\cr #3&#4\cr}}\right)}
\def\vectres#1#2#3{\left(\matrix{#1 \cr #2 \cr #3} \right)}
\def\mattres#1#2#3#4#5#6#7#8#9{\left(\matrix{#1&#2&#3\cr #4&#5&#6\cr
#7&#8&#9\cr} \right)}
\def\Figura#1{{\begin{figure}[h]\caption{#1}\end{figure}}}
\def\jour#1#2#3#4{{\sl #1{}} {\bf #2}, #3\ (#4)}

\begin{center}
{\Large  Wigner functions for curved spaces \\[0.3cm]
	 		I: On hyperboloids} \\[1.5cm]
   {\sc Miguel Angel Alonso},
   {\sc George S.\ Pogosyan},\footnote{Permanent address:
	  Laboratory of Theoretical Physics, Joint Institute for Nuclear
      Research, Dubna, Russia and International Center for Advanced Studies,
   Yerevan State University, Yerevan, Armenia}\\
   and {\sc Kurt Bernardo Wolf}\\[10pt]
   Centro de Ciencias F\'{\i}sicas \\[3pt]
   Universidad Nacional Aut\'onoma de M\'exico \\[3pt]
   Apartado Postal 48--3,  Cuernavaca, Morelos 62251, M\'exico
		\\[1cm]  \today
\end{center}

\vfil

\begin{abstract}

	We propose a Wigner quasiprobability distribution function for
Hamiltonian systems in spaces of constant curvature ---in this paper
on hyperboloids---, which returns the correct marginals and has the
covariance of the Shapiro functions under $\SO{D,1}$ transformations.
To the free systems obeying the Laplace-Beltrami equation on the
hyperboloid, we add a conic-oscillator potential in the hyperbolic
coordinate.  As an example, we analyze the 1-dimensional case on a
hyperbola branch, where this conic-oscillator is the P\"oschl-Teller
potential.  We present the analytical solutions and plot the computed
results.  The standard theory of quantum oscillators is regained in
the contraction limit to the space of zero curvature.

\end{abstract}

\vfil


\eject


\section{Introduction}   \label{sec:1}

	In Hamiltonian systems which a have flat $\Re^D$ configuration
space, among the phase space quasiprobability distribution functions,
the Wigner function \cite{Wigner}, is the only one covariant under
Euclidean translations of phase space \cite{Lee,Feature}.  The
present paper and others that will follow it, aim to the construction
of Wigner functions on configuration spaces that are conic surfaces,
hyperboloids and spheres, which transform under the Lorentz and
rotation groups respectively, and which reproduce the traditional
Wigner function when the conic contracts to the plane.  Hamiltonian
systems on conic manifolds have a natural kinetic energy given by the
Laplace-Beltrami operator, and moreover, on these conics also a
natural oscillator `potential' can be proposed. In one dimension,
this oscillator turns out to be one of the P\"oschl-Teller potentials
\cite{Groposi}.

	In this paper, subtitled I, we propose a Wigner function on the
$D$-dimensional hyperboloid ${\cal H}^D_+$, which generalizes the
ordinary Wigner function on flat phase space.  It displays the
correct marginals, and returns the traditional form of the Wigner
function under In\"on\"u-Wigner contraction to the zero-curvature
limit. The elements and background for this assertion are contained
in Section \ref{sec:2}, including the Shapiro solutions
$\Phi^\ssty{(D)}_{\pp} \of{\xx}$ to the Laplace-Beltrami equation
\cite{Gelfand-Graev-Shapiro}. In Section \ref{sec:3} we present our
proposed definition of Wigner function on the hyperboloid, and verify the
properties of marginality and the contraction limit to flat phase
space. Covariance remains an issue because the Wigner function that
we propose here follows from the covariance of the basis of
wavefunctions, between the argument $\xx$ and the index
$\pp$, as if they were canonically conjugate variables. In this
context, we re-examine the interpretation of momentum coordinates.

	In Section \ref{sec:4} we exemplify the $D$-dim\-ension\-al theory
with a one-dimension\-al {\it sui generis\/} oscillator on one
branch of a hyperbola. This example may appear to be trivial, because
the hyperbola is in most respects equivalent to a straight line.
Nevertheless, the resulting P\"oschl-Teller potential is of
particular interest because the wavefunctions are also the
Clebsch-Gordan (Wigner coupling) coefficients for the
three-dimensional Lorentz algebra, ${\rm so}(2,1) ={\rm
sp}(2,\Re)={\rm su}(1,1)$ \cite{Basu-etal}.  We display
the Wigner functions of some P\"oschl-Teller wavefunctions; these
have not been examined before.  Finally, in Section \ref{sec:5} we
recapitulate the aim and offer the present outlook of our program.


\section{Elements of phase space and hyperboloids}  \label{sec:2}

	In his fundamental article \cite{Wigner}, Wigner proposed a
distribution function to represent on phase space the wavefunctions
of pure and of mixed states in quantum systems. In this Section we
recall the definition and properties that we shall generalize from
flat to conic spaces, using the Laplace-Beltrami operator and the
Shapiro functions.

\subsection{Wigner function on flat phase space}

	In $D$-dimensional flat configuration space $\xx\in\Re^D$, the
generalized Dirac basis of plane waves solves the Helmholtz equation
\be
		{-}\!\Delta\,\phi\of\xx=p^2\,\phi\of\xx,\qquad
	\phi_{\bf p}\of{\bf x} =
					\exp(i\pp\cdot\xx), \quad \pp\in\Re^D.
							\label{eq:basis-of-plane-waves}
\ee
When we write $p=+(\pp\cdot\pp)^{1/2}$, $\nn=\pp/p$, and call
$\pp=p\,\nn$ the momentum or wavenumber vector, the functions
$\phi_{\bf p}\of{\bf x}$ represent plane waves in the direction
of the unit vector $\nn\in{\cal{S}}_{D-1}$ in the
$(D{-}1)$-dimensional sphere manifold.
In the quantum model with natural units $\hbar=1$, $p$ has units of
inverse length; in the wave optical model, $p$ is the
wavenumber of light.

	The basis of plane wave functions (\ref{eq:basis-of-plane-waves})
plays many roles: it  provides the Fourier transform kernel
which bridges the  configuration and momentum realizations,
it constitutes a basis for representations of the Euclidean group,
and it serves for the construction of the $\Re^{2D}$-Wigner function
of wavefields $f\of\xx$, $g\of\xx$ through the equivalent expressions
\bea
	W_{\Re^D}(f,g|\xx,\pp)&=&\frac1{(2\pi)^D}
		\int_{\Re^D}\!\!d^D\zz\,
			f\of{\xx{-}\onehalf\zz}^*\,
				e^{-i\pp\cdot\zz}\,
					g\of{\xx{+}\onehalf\zz}
									\label{eq:Wig-traditional}\\
		&=&\frac1{(2\pi)^D}\int_{\Re^D}\!\!\!d^D\zz\,
		f\of{\xx{-}\onehalf\zz}^*\,e^{+i\pp\cdot(\xx{-}\frac12\zz)}\,
		e^{-i\pp\cdot(\xx{+}\frac12\zz)}\,g\of{\xx{+}\onehalf\zz}
										\nonumber\\
			&=&\frac1{(2\pi)^D}\int_{\Re^D}\!\!\!d^D\xx'
					\!\int_{\Re^D}\!\!\!d^D\xx''\,
			f\of{\xx'}^*\,g\of{\xx''} \nonumber\\
		& &{}\qquad\qquad\qquad{}\times
			\phi_\pp\of{\xx'}\,
				\delta^D(\xx{-}\onehalf\of{\xx'{+}\xx''})\,
			\phi_\pp\of{\xx''}^*. \label{eq:Wig-delta}
		\eea
This has the well-known properties of being sesquilinear in the
functions, real for $f=g$, with the marginal projections
$\int_{\Re^D}d\pp\,W=f\of\xx^*\,g\of\xx$,
$\int_{\Re^D}d\xx\,W=\widetilde f\of\pp^*\,\widetilde g\of\pp$ (the tilde
indicates ordinary Fourier transformation, $F:f=\widetilde f$), and
covariant under translations in coordinate and momentum spaces
\bea
	T_{\bf a}:f\of\xx=f\of{\xx{-}{\bf a}}
		&\Rightarrow&
			W_{\Re^D}(T_{\bf a}{:}f,T_{\bf a}{:}g|\xx,\pp)
				= W_{\Re^D}(f,g|\xx-{\bf a},\pp),
								\label{eq:covar-x-a}\\
	\widetilde T_{\bf b}:f\of\xx=e^{i{\bf b}\cdot\xx}f\of\xx
		&\Rightarrow&
		W_{\Re^D}(\widetilde T_{\bf b}{:}f,\widetilde T_{\bf b}{:}g|\xx,\pp)
				= W_{\Re^D}(f,g|\xx,\pp-{\bf b}),
								\label{eq:covar-p-b}\\
	F:f\of\xx=\widetilde f\of\xx
		&\Rightarrow&
		W_{\Re^D}(F{:}f,\ F{:}g\,|\xx,\pp)
				= W_{\Re^D}(f,g\,|\pp,-\xx).
								\label{eq:covar-p-F}
\eea
The last intertwining by the Fourier transform was known when
Garc\'{\i}a-Calder\'on and Moshinsky noticed that the Wigner 
function is covariant also under the larger group of $\Sp{2D}$ linear
canonical transformations of phase space \cite{G-C-M}. This is
exceptional in the sense that the Heisenberg-Weyl algebra [whose 
generators are the phase space translations
(\ref{eq:covar-x-a})--(\ref{eq:covar-p-b}) ---and the unit that 
generates a commuting phase factor] has the outer automorphism group
$\Sp{2D}$. This accident does not occur for Lorentz algebras, so we
should not expect similar covariances of the Wigner function under 
groups larger than $\SO{D,1}$.

\subsection{Laplace-Beltrami operator on the hyperboloid}

	The purpose of this paper is to generalize the expression of the
Wigner function (\ref{eq:Wig-traditional})--(\ref{eq:Wig-delta}) with
functions on a $D$-dimensional space $\xx\in\Re^D$ of constant
curvature. This manifold can be seen in an `ambient' space of
$D{+}1$ dimensions as a hyperboloid, with vectors
$x=(x_0,\xx)\in\Re^{D+1}$. 

	Consider the upper sheet of the two-sheeted hyperboloid
${\cal H}^D_+\subset\Re^{D+1}$ of hyperbolic radius $R>0$,
\be
		\abs{x}^2 = x_0^2 - \xx^2 = R^2, \qquad
					  \xx^2= x_1^2 + x_2^2 +\cdots+ x_D^2.
						\label{eq:Hyper1}
\ee
In this ambient Minkowski space, the isometry group is the Poincar\'e
group ${\rm ISO}(D,1)_+^\uparrow$, in place of the Euclidean group
${\rm ISO}(D)_+$ of flat space. The Lie algebra $\so{D,1}$ has then
the standard realization 
\be
		M_{j,k} =  x_j \partial_{x_k} - x_k \partial_{x_j},
			\quad
		M_{0,k} = x_0 \partial_{x_k} + x_k \partial_{x_0},
						\qquad j,k=1,2,\ldots,D.
								\label{eq:soN1-generators}
\ee
The second-order Casimir operator, $\cal C$, which
is an invariant under the group $\SO{D,1}_+^\uparrow$, is
($-R^2$ times) the Laplace-Beltrami operator on ${\cal H}^D_+$,
namely
\be
	\frac1{R^2}{\cal C}=-\Delta_\ssr{LB} = \frac1{R^2}\bigg(
			\sum_{1\le j<k\le D} M_{j,k}^2
			-\sum_{1\le k\le D} M_{0,k}^2\bigg).
								\label{eq:Casimir}
\ee
In the unitary irreducible representation spaces of the $D$-dimensional
Lorentz group belonging to the most degenerate continuous series
indicated by $p$ \cite{Gelfand}, the operator (\ref{eq:Casimir}) has
a real lower-bound spectrum, as does (\ref{eq:basis-of-plane-waves}).
The wavefunctions of the free system on the hyperboloid are the
solutions to the equation
\be
	{\displaystyle\Delta_\ssr{LB} \,f\of\xx
	= - \left[\bigg(\frac{D-1}{2R}\bigg)^2 + p^2\right]\,f\of\xx
	= -\frac{ \lambda  ( \lambda +D-1)}{R^2} \,f\of\xx, \atop
		p\in\Re^+_0=[0,\infty),	\qquad
				 \lambda =-\onehalf(D-1) - i p R.}
								\label{eq:Laplace-Beltrami-D}
\ee
Any wavefield of a given wavenumber $p$ is a solution of this
equation. 

\subsection{Shapiro functions}

	A privileged basis for the solutions of the Laplace-Beltrami
equation (\ref{eq:Laplace-Beltrami-D}) were given by Gel'fand, Graev
and Shapiro \cite{Gelfand-Graev-Shapiro} in the form of $D$-dimensional
plane waves of momentum $\pp=p\,\nn$, with positive wavenumber $p$
and in the direction of a unit vector on the sphere
$\nn\in{\cal S}^{D-1}$,
\be
	\begin{array}{rcll}
	\Phi^\ssty{(D)}_{\pp} \of{x}
		&=&\displaystyle
		\left(\frac{x_0-\nn\cdot\xx}{R}\right)^{-\frac12(D-1)-ipR}& \\
		&=&\displaystyle(\cosh\chi- \nn\cdot{\pmbf\xi}\sinh\chi
					)^{-\frac12(D-1) - ipR}, &
								\end{array}	\label{eq:Shapiro-functions}
\ee
where functions $f\of{x}$ on the hyperboloid $x\in{\cal H}^D_+$
($x^2=R^2$) will be denoted, according to convenience, by
\be
	f\of{x}=f(x_0,\xx)=f\of\xx,\quad
		\begin{array}{l} x_0=+\sqrt{R^2+\xx^2}=R\cosh\chi\ge R,\\
			\xx=R\,{\pmbf\xi}\sinh\chi\in\Re^D,\ \chi\in\Re_0^+,\
						{\pmbf\xi}\in{\cal S}^{D-1}. \end{array}
							\label{eq:f-on-hyperboloid}
\ee

	The {\it Shapiro\/} functions (\ref{eq:Shapiro-functions}) are a
Dirac basis for functions on the hyperboloid, which are orthogonal
and complete over \xx- and \pp-spaces:
\bea
	\frac{R}{(2\pi)^D}
	\int_{\xx\in\Re^D} {d^D\xx\over x_0}\,
		\Phi^\ssty{(D)}_\pp \of{x}^*\,\Phi^\ssty{(D)}_{\pp'} \of{x}
	&=&N^\ssty{(D)}(p)\,\delta^D(\pp-\pp'),\label{eq:Phi-orthogonality}\\
	\frac{1}{(2\pi)^D}
	\int_{\pp\in\Re^D} {d^D\pp\over N^\ssty{(D)}(p)}\,
		\Phi^\ssty{(D)}_\pp\of{x}^*\,\Phi^\ssty{(D)}_\pp\of{x'}
		&=&\delta^D(x,x'),	\label{eq:Phi-completeness}
\eea
with the measure and Dirac $\delta$ under $\int_{{\cal H}^D_+}d^Dx
=R\int_{\Re^D}d^D\xx/x_0$,
\bea
	N^\ssty{(D)}(p)&=&\left|\frac{\Gamma(ipR)}{
			\Gamma\Big(\onehalf(D-1)+ipR\Big)}\right|^2
					\,(pR)^{D-1}, 		\label{eq:N(p)}\\
	\delta^D(x,x')&=&\frac{x_0}R\, \delta^D(\xx-\xx')
					=\sqrt{1+\frac{\xx^2}{R^2}}\,\,\delta^D(\xx-\xx').
							\label{eq:Delta-D}
\eea
In particular, $N^\ssty{(1)}(p)=1$,
$N^\ssty{(2)}(p)=\coth(pR)$, and $N^\ssty{(3)}(p)=1$.

	The In\"onu-Wigner contraction limit of the Lorentz to the
Euclidean group $\SO{D,1}_+^\uparrow\to{\rm ISO}(D)_+$ is the limit
$R\to\infty$ in our expressions for vectors with $x_0\approx R$,
$\xx^2\ll R^2$, and $\pp=p\,{\nn}$ as before, \ie,
\bea
	\lim_{R\to\infty}\Phi^\ssty{(D)}_{\pp} \of{x}
	&=&\lim_{R\to\infty}
	\left(\frac{x_0-\xx\cdot\nn}{R}\right)^{-\frac12(D-1)-ipR}
								\nonumber\\
	&\approx&\lim_{R\to\infty}\left(1-\frac{\xx\cdot\nn}{R}\right)^{-ipR}
						=\exp(i\,\xx\cdot\pp).
								\label{eq:contraction-Phi}
\eea
Correspondingly, $\lim_{R\to\infty}N^\ssty{(D)}\of{p}=1$ and
$\delta^D(x,x')\to\delta^D(\xx-\xx')$.

\subsection{Momentum space for the hyperboloid}

	 The Shapiro functions $\{\Phi^\ssty{(D)}_{\pp}
\of{\xx}\}_{\pp\in\Re^D}$ in  (\ref{eq:Shapiro-functions}) serve as
the integral transform kernel between functions of \xx\ on the hyperboloid,
$f\of\xx$, and conjugate functions of \pp, that has the
interpretation of momentum or wavenumber space, and is indicated
$\widetilde f\of\pp$. Using (\ref{eq:f-on-hyperboloid}) for
$\xx,\pp\in\Re^D$, one writes
\bea
	\widetilde f\of\pp &=& \frac{R}{(2\pi)^{D/2}}\int_{\xx\in\Re^D}
				\frac{d^D\xx}{x_0}\Phi^\ssty{(D)}_\pp \of{\xx}^*\,
				f\of\xx,   \label{eq:Shapiro-Mellin-tfm}\\
	f\of\xx &=& \frac1{(2\pi)^{D/2}}\int_{\pp\in\Re^D}
			\frac{d^D\pp}{N^\ssty{(D)}(p)}\Phi^\ssty{(D)}_\pp\of{\xx}\,
				\widetilde f\of\pp.  \label{eq:Shapiro-Mellin-inv}
\eea
This {\it Shapiro transform\/} has been used as 
a relativistic analogue 
of the Fourier transform (the physical context here, though, is {\bf
not} that of space-time relativity, as we shall clarify below), and
is a vector form of one of the two branches of the bilateral Mellin
transform \cite{KBW-book}. Here the Shapiro transform 
replaces the traditional Fourier transform in the
definition of a momentum space $\pp\in\Re^D$, canonically
conjugate with respect to this basis, to a configuration space of
constant curvature. The corresponding Parseval relation is 
\be
	R\int_{\xx\in\Re^D}\!\frac{d^D\xx}{x_0}\,f\of\xx^*\,g\of\xx
	= (f,\,g)_{{\cal H}^D_+}
	=\int_{\pp\in\Re^D}\frac{d^D\pp}{N^\ssty{(D)}(p)}\,
				\widetilde f\of\pp^*\,\widetilde g\of\pp.
							\label{eq:Shapiro-Mellin-Parseval}
\ee

	The manifold of momentum $\pp=p{\bf n}\in\Re^D$ ($p\in\Re^+$
and ${\bf n}\in{\cal S}^{D-1}$), can be placed also in a
$(D+1)$-dimensional `ambient' space, where it occupies the cone
$\varpi=(p,\pp)\in\bigvee^+$. The momentum thus defined by the
Shapiro functions has certain features however, which do not
correspond to those of a standard relativistic momentum vector. If 
$f\of\xx$ is a monochromatic wavefield with a definite
value of $p$, this wavenumber will not change under $\SO{D,1}$
translations of the hyperboloid (`boosts'), because it is the
invariant value of the Casimir operator
(\ref{eq:Casimir})--(\ref{eq:Laplace-Beltrami-D}).  
Only the direction of momentum, $\bf n$, can shift over the sphere;
it will do so following the well-known Bargmann deformation of the circle
\cite{Bargmann}, where the colatitude angle `boosts' as 
$\tan\onehalf\phi\mapsto e^{-\zeta}\tan\onehalf\phi$ for rapidity
$\zeta\in\Re$. Quotation marks are used for `boost' because here we
mean a translation in the hyperboloid, and not the well-known
relativistic acceleration.   


\subsection{Oscillators on conics}

	The Laplace-Beltrami equation
(\ref{eq:Casimir})--(\ref{eq:Laplace-Beltrami-D}) provides the free
fields (whose ener\-gy is purely kinetic) on the hyperboloid. One
introduces a potential energy term, as in the Schr\"odinger equations
of quantum mechanics, by adding a function of position $V\of\xx$ 
\cite{Groposi,Higgs-Leemon,schrodinger},
\be
	\Big(\frac{-1}{2\mu}\Delta_\ssr{LB} + R^2\,V\of\xx\Big)f\of\xx 
					=R^2 E f\of\xx. 	\label{eq:Schrodinger-osc-D}
\ee
In quantum mechanics $\mu=m/\hbar^2$, where $m$ is the particle mass.
For application in paraxial wave optics, we recall the interpretation
where the extra term characterizes the refractive index anomaly of
the medium,
\be
	n\of\xx=n_\circ-\nu\of\xx,\quad n_\circ=n(0), \qquad 
		\begin{array}{rcl}
		n_\circ&\leftrightarrow&\mu,\\
		\nu\of\xx&\leftrightarrow& V\of\xx.\end{array}
									\label{eq:nyV}
\ee

	A straightforward and useful generalization of the
$\SO{D}$-isotropic harmonic oscillator potential from flat to conic
$D$-dimensional configuration space is \cite{Higgs-Leemon}
\be
	V\of\xx=\onehalf\mu\omega^2R^2 \frac{\abs\xx^2}{x_0^2}
	  =\onehalf\mu\omega^2R^2 \tanh^2 \chi 
	  =\onehalf\mu\omega^2R^2\, (1-\hbox{sech}^2 \chi),
					\label{eq:conicHO--potential}
\ee
where $\chi\in\Re_0^+$ is the hyperbolic angle coordinate defined in
Eqs.\ (\ref{eq:f-on-hyperboloid}). This is the P\"oschl-Teller
`secant-hyperbolic-squared' trough.


\section{Wigner function on the hyperboloid}   \label{sec:3}

	With the Shapiro basis of wavefunctions of the free system,
we construct now our proposed Wigner function following the
double-integral form in Eq.\ (\ref{eq:Wig-delta}) for two
wavefunctions, $f\of{x}$ and $g\of{x}$, by means of integrals on two
hyperboloids, $\abs{x'}=R$ and $\abs{x''}=R$.

\subsection{Definition}

		With the measures in Eqs.\ (\ref{eq:Phi-orthogonality}) and
the Shapiro functions in (\ref{eq:Shapiro-functions}), we define
the Wigner function on the hyperboloid by 
\be \begin{array}{rcl}
	W_{\cal H}(f,g|\xx,\pp)
	&=&\displaystyle\frac{R^2}{(2\pi)^D}
	\int_{\xx'\in\Re^D}\!\! {d^D\xx'\over x_0'}
	\int_{\xx''\in\Re^D}\!\! {d^D\xx''\over x_0''}\,\,
				 f\of{x'}^*\,g\of{x''}\\[10pt]
		& &\displaystyle\qquad\quad{}\times\Phi^\ssty{(D)}_\pp\of{x'}
				\,{\it\Delta}^D(x;x',x'')\,
			\Phi^\ssty{(D)}_\pp\of{x''}^*,\end{array}
				\label{eq:Wig-new}
\ee
where ${\it\Delta}^D(x;x',x'')$ takes the place of the Dirac delta
$\delta^D(\xx{-}\onehalf\of{\xx'{+}\xx''})$ on flat space, 
Eq.\ (\ref{eq:Wig-delta}), and which will be detailed below.

	The crucial property that we must require of this
`binding-$\it\Delta$' in (\ref{eq:Wig-new}) is that it should guarantee
that $x$ be the midpoint of the {\it geodesic\/} between $x'$ and
$x''$, so that all three points lie on the hyperboloid ${\cal
H}^D_+$.  We achieve this in the following way \cite{Alonso}: given
$x\in\Re^{D+1}$ in the upper sheet of a two-sheeted hyperboloid, 
we build any $y\in\Re^{D+1}$ on a one-sheeted hyperboloid 
$\widetilde{\cal H}^D$ of the same radius $R$, such that
it be Minkowski-orthogonal to $x$,
\be
	y=(y_0,\yy),\quad
	\abs{y}^2=y_0^2-\yy^2=-R^2,\qquad
	x_0y_0-\xx\cdot\yy=0.
				\label{eq:Mikowski-orthogonal}
\ee
Then, we can express $x'$ and $x''$ as vectors obtained from $x$ and
$y$ as follows:
\be
	x'=x\cosh\onehalf\tau-y\sinh\onehalf\tau,\qquad
	x''=x\cosh\onehalf\tau+y\sinh\onehalf\tau,
									\label{eq:boost-by-beta}
\ee
where $x',x''\in{\cal H}^D_+$ for all $\tau\in\Re$. Also, it is easy
to show that 
\be
	x_0'x_0''-\xx'\cdot\xx''=R^2\cosh\tau,\quad
	x_0x_0'-\xx\cdot\xx'=x_0x_0''-\xx\cdot\xx''=R^2\cosh\onehalf\tau,
							\label{eq:xxprime-in-H}
\ee
\ie, the geodesic distance between $x'$ and $x''$ is $R\tau$,
while $x$ is at $\onehalf R\tau$ from both $x'$ and $x''$.
The arguments $x'$ and $x''$ in the expression (\ref{eq:Wig-new})
thus emulate the arguments $\xx\pm\onehalf\zz$ in
(\ref{eq:Wig-traditional}) with the parameter $\onehalf\tau$.

	Using (\ref{eq:Delta-D}) and the parameter $\tau$ in
(\ref{eq:xxprime-in-H}), we propose the binding-$\it\Delta$
in (\ref{eq:Wig-new}) to be
\be
	{\it\Delta}^D(x;x',x'')
		= \frac{x_0}R\,
		\delta^D\Big(\xx-\frac{\xx'+\xx''}{2\cosh\onehalf\tau}\Big).
					\label{eq:def-capital-Delta}
\ee
This will yield the correct marginals (to be seen below) due to its
properties 
\be
	 {\it\Delta}^D(x;x',x')=\frac{x_0}R\,\delta^D(\xx-\xx'),\qquad
	R\int_{\xx\in\Re^D} {d^D\xx\over x_0}\,{\it\Delta}^D(x;x',x'')=1.
				\label{eq:inte-Delta-D}
\ee

\subsection{Integral forms}

	 The $2D$-fold integral form of the Wigner function in
(\ref{eq:Wig-new}) contains Dirac $\delta$'s; it can therefore be
brought to a $(D+1)$-fold integral noting that the definition of
$y\in\widetilde{\cal H}^D$ leaves the freedom of rotating \yy\ around
\xx\ on a sphere ${\cal S}^{D-1}$. When we change variables from $x'$
and $x''$ to $x$ and $y$ according to (\ref{eq:boost-by-beta}), we
reduce the integration to $\yy$ and $\tau$ while keeping
Minkowski-orthogonality. The proposed Wigner function
(\ref{eq:Wig-new}) then becomes 
\bea
	W_{\cal H}(f,g|\xx,\pp)&=& \frac{R^2}{(2\pi)^D}
	\int_0^\infty (\sinh\tau)^{D-1} \,d\tau
	\int_{y\in\widetilde{\cal H}^D} {d^D \yy}\,
		\delta(x_0y_0-\xx\cdot\yy)
						\nonumber\\ & &
		\times f\of{x\cosh\onehalf\tau{-}y\sinh\onehalf\tau}^*\,
				   g\of{x\cosh\onehalf\tau{+}y\sinh\onehalf\tau}
						\qquad{}\label{eq:newest-Wigner}\\ & & \times
		\Phi^\ssty{(D)}_\pp\of{x\cosh\onehalf\tau{-}y\sinh\onehalf\tau}
		\Phi^\ssty{(D)}_\pp\of{x\cosh\onehalf\tau{+}y\sinh\onehalf\tau}^*.
						\nonumber
\eea

	The Dirac $\delta$ remaining in (\ref{eq:newest-Wigner}) can be
used to find a third alternative form of the Wigner function. This is
obtained with the parametrization of the ambient-space vectors given
by 
\be
	x=(x_0,\xx)=R(\cosh\chi,{\pmbf\xi}\sinh\chi),\quad
	y=(y_0,\yy)=R(\sinh\omega,{\pmbf\eta}\cosh\omega),
		\label{eq:x-y-etc}
\ee
where $\pmbf\xi$ and $\pmbf\eta$ are unit vectors on the sphere
${\cal S}^{D-1}$ and $\chi,\omega\in\Re_0^+$.  The Dirac $\delta$ in
Eq.\  (\ref{eq:newest-Wigner}) is then
\be
	\delta(x_0y_0-\xx\cdot\yy)
	= \frac1{R^2}
	{\cosh\Omega\over\cosh\chi}\,\delta(\omega-\Omega),
				\quad\hbox{with}\quad
	\tanh\Omega={\pmbf\xi}\cdot{\pmbf\eta}\,\tanh\chi.
						\label{eq:DDeltass}
\ee
The differential $d^D\yy$ of the integral in
$y\in\widetilde{\cal H}_+^D$ becomes
$R^D(\cosh\omega)^{D-1}\,\*d\omega\,\*d^{D-1}{\pmbf\eta}$, so the
Wigner function (\ref{eq:Wig-new}) becomes a $D$-fold integral with
the structure of (\ref{eq:Wig-traditional}), {\it viz.},
\bea
	W_{\cal H}(f,g|\xx,\pp)&=&
	\frac1{(2\pi)^D}\int_0^\infty (\sinh\tau)^{D-1} \,d\tau
	\int_{{\cal S}^{D-1}} {\abs\yy^D\over\cosh\chi}\,
					{d^{D-1}{\pmbf\eta}}	\nonumber\\ & &
		\times f\of{x\cosh\onehalf\tau{-}y\sinh\onehalf\tau}^*\,
				   g\of{x\cosh\onehalf\tau{+}y\sinh\onehalf\tau}
						\label{eq:newestestest-Wigner}\\ & & \times
		\Phi^\ssty{(D)}_\pp\of{x\cosh\onehalf\tau{-}y\sinh\onehalf\tau}
		\Phi^\ssty{(D)}_\pp\of{x\cosh\onehalf\tau{+}y\sinh\onehalf\tau}^*,
							\qquad{}\nonumber
\eea
with 
\be	 y= R(\sinh\Omega,\ {\pmbf\eta}\cosh\Omega)
			= R{({\pmbf\xi}\cdot{\pmbf\eta}\tanh\chi,\ {\pmbf\eta})
				\over\sqrt{1-({\pmbf\xi}\cdot{\pmbf\eta}\tanh\chi)^2}}.
						\label{eq:define-y}
\ee

\subsection{Marginal projections}

	The marginal projections obtained by integrating the proposed Wigner
function (\ref{eq:Wig-new}) over momentum and configuration space
should yield, respectively, $f\of\xx^*g\of\xx$ and $\widetilde
f\of\pp^*\widetilde g\of\pp$ as defined in
(\ref{eq:Shapiro-Mellin-tfm})--(\ref{eq:Shapiro-Mellin-inv}).  The
two marginals follow from the orthogonality and
completeness relations of the Shapiro functions, Eqs.\
(\ref{eq:Phi-orthogonality})--(\ref{eq:Phi-completeness}).

	The integration of the Wigner function over $\Re^D$ momentum
space with the measure $1/N^\ssty{(D)}\of{p}$ in (\ref{eq:N(p)}), is
\bea
	M_{\cal H}(f,g|\xx)
	&=& \int_{\pp\in\Re^D} {d^D\pp\over N^\ssty{(D)}(p)}\,
		W_{\cal H}(f,g|\xx,\pp) \nonumber\\
	&=&\!\!R^2 \int_{\xx'\in\Re^D}\!\! {d^D\xx'\over x_0'}
	\int_{\xx''\in\Re^D}\!\! {d^D\xx''\over x_0''}\,\,
				f\of{x'}^*\,g\of{x''}\,
		{\it\Delta}^D(x;x',x'')\,\delta^D\of{x',x''}\nonumber\\
	&=& R\int_{\xx'\in\Re^D}\!\! {d^D\xx'\over x_0'}
				f\of{x'}^*\,g\of{x'}\,
			{\it\Delta}^D(x;x',x')\nonumber\\
	&=& f\of{\xx}^*\,g\of{\xx},   \label{eq:marginal-in-x}
\eea
where we used (\ref{eq:Phi-completeness}), (\ref{eq:Delta-D}),
and the first property of the binding-$\it\Delta$ in
(\ref{eq:inte-Delta-D}).

	Similarly, the integration over $\Re^D$ configuration space with
the measure $R/x_0$ in (\ref{eq:Delta-D}), is

\vbox{\bea
	M_{\cal H}(f,g|\pp)
	&=& R\int_{\xx\in\Re^D}
		{d^D\xx\over\sqrt{R^2+\xx^2}}\,
			W_{\cal H}(f,g|\xx,\pp) \nonumber\\
	&=&\frac{R^2}{(2\pi)^D} \int_{\xx'\in\Re^D}\!\! {d^D\xx'\over x_0'}\,
			f\of{x'}^*\,\Phi^\ssty{(D)}_{\pp}\of{x'}\,
	\int_{\xx''\in\Re^D}\!\! {d^D\xx''\over x_0''}\,
			g\of{x''}\,\Phi^\ssty{(D)}_{\pp}\of{x''}^*\nonumber\\
	&=& \widetilde f\of{\pp}^*\,\widetilde g\of{\pp},
							\label{eq:marginal-in-p}
\eea}

\noindent where we used the second property of the binding-$\it\Delta$ in
(\ref{eq:inte-Delta-D}) and the Shapiro transform
(\ref{eq:Shapiro-Mellin-tfm}).

	Finally, integrating over the whole of phase space with the
appropriate measures in the Parseval relation
(\ref{eq:Shapiro-Mellin-Parseval}), we have the total probability
\be
	R\int_{\xx\in\Re^D}\!\frac{d^D\xx}{x_0}\,M_{\cal H}(f,g|\xx)
	= (f,\,g)_{{\cal H}^D_+}
	=\int_{\pp\in\Re^D}\frac{d^D\pp}{N^\ssty{(D)}(p)}\,
				M_{\cal H}(f,g|\pp).
							\label{eq:Wigner-Parseval}
\ee


\subsection{Covariance under rotations and conic translations}

	Under rotations ${\bf R}\in\SO{D}$, wavefunctions $f\of{x_0,\xx}$
transform through
\be
	T({\bf R}):f\of{x}=f(x_0,{\bf R}^{-1}\xx).  \label{eq:act-rot}
\ee
In particular, the basis of Shapiro functions
(\ref{eq:Shapiro-functions}) transforms as
\be
	T({\bf R}):\Phi^\ssty{(D)}_{p\,\nn} (x_0,\xx)
		= \Phi^\ssty{(D)}_{p\,\nn} (x_0,{\bf R}^{-1}\,\xx)
		= \Phi^\ssty{(D)}_{p\,{\bf R}\nn} (x_0,\xx).
							\label{eq:covar-rot}
\ee
Applying rotations $T({\bf R})$ to the wavefields $f$ and $g$ in the
Wigner function (\ref{eq:Wig-new}), we next change variables to
$\xx'={\bf R}\bar\xx'$ and $\xx''={\bf R}\bar\xx''$ (the
ambient $x_0$-components behave as scalars), then use
(\ref{eq:covar-rot}) for ${\bf R}^{-1}$, noting that the
binding-$\it\Delta$ in (\ref{eq:def-capital-Delta}) is invariant,
${\it\Delta}^D(\bar x;\bar x',\bar x'')={\it\Delta}^D(x;x',x'')$ for 
$\bar\xx={\bf R}^{-1}\xx$, and so are the measures
$d^D\xx'=d^D\bar\xx'$. It thus follows that the Wigner function
(\ref{eq:Wig-new}) is covariant under rotations, fulfilling 
\be
	W_{\cal H}(T({\bf R}){:}f,T({\bf R}){:}g|\xx,\pp)
	= W_{\cal H}(f,g|{\bf R}^{-1}\xx,{\bf R}^{-1}\pp).
						\label{eq:rotation-covariance}
\ee

	Now consider translations by $\zeta$ (`boosts' of rapidity $\zeta$)
${\bf B}_{\bf m}(\zeta)\in\SO{D,1}_+^\uparrow$ in the direction of unit 
${\bf m}\in{\cal S}^{D-1}$, which transform the ambient space
vectors preserving the constant-curvature subspaces 
$x\in{\cal H}^D_+$ for each radius $R>0$.
We denote by $\xx_{\parallel{\bf m}}$ and $\xx_{\perp{\bf m}}$
the projections of \xx\ parallel and perpendicular to the direction
of {\bf m}, so that $\xx=\xx_{\parallel{\bf m}}+\xx_{\perp{\bf m}}$.
Then, wavefunctions on the hyperboloid transform as
\be
	T({\bf B}_{\bf m}(\zeta)):f\vectres{x_0}{
		\xx_{\parallel{\bf m}}}{\xx_{\perp{\bf m}}}
	= f\vectres{x_0\cosh\zeta-{\bf m}\cdot\xx\,\sinh\zeta}{
		\xx_{\parallel{\bf m}}\cosh\zeta-x_0\,{\bf m}\sinh\zeta}{ 
						\xx_{\perp{\bf m}}}.
								\label{eq:act-boost}
\ee
When this transformation is applied to the plane-wave basis of
Shapiro functions, their directions ${\bf n}$ on the sphere change,
and they acquire a multiplier factor: 
\be
	T({\bf B}_{\bf m}(\zeta)):\Phi^\ssty{(D)}_{p\,\nn} (x_0,\xx)
		= (\cosh\zeta{+}{\bf m{\cdot}\nn}\,\sinh\zeta)^{-\frac12(D-1)-ipR}
			\,\Phi^\ssty{(D)}_{p\,\nn'}(x_0,\xx),
							\label{eq:covar-boost}
\ee
where the components of ${\bf n}'\in{\cal S}^{D-1}$ that are
orthogonal and parallel to $\bf m$ are 
\be
	\nn'_{\perp{\bf m}}
		= \frac{\nn_{\perp{\bf m}}}{
				\cosh\zeta+\nn\cdot{\bf m}\sinh\zeta},\quad
	\nn'_{\parallel{\bf m}}
		= \frac{\nn\cdot{\bf m}\cosh\zeta+\sinh\zeta
			  }{\cosh\zeta+\nn\cdot{\bf m}\sinh\zeta},
							\label{eq:nnprime-perp-parallel}
\ee
within the same $\SO{D,1}_+^\uparrow$ irreducible representation
characterized by the invariant wavenumber, $p$. If the
angle from {\bf m} to \nn\ is $\phi$, it will transform through the
well-known Bargmann \SO{2,1} map of the circle.

	 The expression in the multiplier factor of Eq.\
(\ref{eq:covar-boost}),  
\be
	\mu({\bf m},\zeta;\nn)=\cosh\zeta{+}{\bf m{\cdot}\nn}\,\sinh\zeta,
										\label{eq:multiplier}
\ee
is, not coincidentally, $p'/p$ ---if the ($D{+}1$)-vector
$\varpi=(p,\pp)\in\bigvee^+$ were allowed to transform as a
`lightlike' vector in relativity, \ie, without being constrained to
its $p$-sphere. Under the inner product (\ref{eq:Phi-completeness}),
the $\SO{D,1}$ boost with the multiplier (\ref{eq:covar-boost}) is
unitary nonetheless, because the measure in \pp-space 
is $d^D\pp=p^D\,dp\,d^{N-1}\nn$, and while $p$ is invariant, from
(\ref{eq:nnprime-perp-parallel}) it follows that 
$d^{D-1}\nn=\mu({\bf m},\zeta;\nn)^{D-1}\,d^{D-1}\nn'$. This cancels
the absolute square of the multiplier (\ref{eq:multiplier}) in the
Shapiro functions (\ref{eq:act-boost}). This type of covariance
modulo a multiplier function is determined by the Shapiro function
basis; we may call Eqs.\ (\ref{eq:act-boost}) and
(\ref{eq:covar-boost}) the {\it Shapiro covariance\/} between the
conjugate transformations in \xx\ and \pp.

	When the wavefields in the Wigner function (\ref{eq:Wig-new}) are
translated within the hyperboloid by (\ref{eq:act-boost}), the
ambient vectors $x$ are multiplied by a $(D{+}1)\times(D{+}1)$
(`boost') matrix ${\bf B}_{\bf m}(\zeta)$, to $x'\mapsto\bar x'
={\bf B}_{\bf m}(\zeta)^{-1}\,x'$ and  
$x''\mapsto\bar x''={\bf B}_{\bf m}(\zeta)^{-1}\,x''$. Under this
transformation, the measures are again invariant,
$d^D\xx'/x'_0=d^D\bar\xx'/\bar x'_0$, etc., and so is
$\Delta^D(x;x',x'') =\Delta^D(\bar x;\ \bar x',\bar x'')$; hence,
$\bar x ={\bf B}_{\bf m}(\zeta)^{-1}\,x$ will appear in the first
argument of the transformed Wigner function. But the corresponding
transformation of \pp\ in each of the two Shapiro functions, Eqs.\
(\ref{eq:covar-boost}) and (\ref{eq:multiplier}), yields a multiplier
factor. The imaginary exponents of $\mu({\bf m},\zeta,{\bf n})$
cancel, and there remains a positive net multiplier factor:
\bea
	& &{}\hskip-20pt W_{\cal H}\Big(
	T[{\bf B}_{\bf m}(\zeta)]{:}f,\,T[{\bf B}_{\bf m}(\zeta)]{:}g
						\Big|\xx,p\,\nn\Big)
							\nonumber\\
	& &{}= \Big(\mu({\bf m},\zeta,{\bf n})\Big)^{-D{+}1}\,
		W_{\cal H}\Big(f,g
	\Big|{\bf B}(\zeta)^{-1}{:}\xx,\ 
					p\,{\bf B}_{\bf m}(\zeta)^{-1}{:}\nn\Big),
						\label{eq:non-rel-covariance}
\eea
where $\nn'={\bf B}_{\bf m}(\zeta)^{-1}{:}\nn$ is given by
(\ref{eq:nnprime-perp-parallel}). Note that in the $D=1$-dimensional
case, the multiplier factor is 1.

	Covariance of the Wigner function is usually understood in the
simple form it has under rotations, as given by
(\ref{eq:rotation-covariance}). Under these transformations, the
hyperboloid in the ambient $x$-space rotates on its axis, and in the
momentum plane the circles \nn\ of all radii $p$ rotate in sinchrony.
Translations within the hyperboloid (\ref{eq:non-rel-covariance}) on
the other hand, deform the ambient and projected space vectors, $x$
and $\xx$, through (\ref{eq:act-boost}); momentum space is
concurrently squeezed in the direction of the translation so that its
points move on constant-$p$ circles and with a common Bargmann
deformation of the angle. Since areas are not conserved in momentum
space, a multiplier function of \pp\ is necessary for the Wigner
function to ensure the total conservation of probability
(\ref{eq:Wigner-Parseval}).

\subsection{Contraction limit}

	We now show that, when $f\of{x}$ and $g\of{x}$ are significantly
different from zero only within a small, essentially flat patch of
the hyperboloid, the definition of the Wigner function in Eq.\
(\ref{eq:newestestest-Wigner}) reduces to the standard Wigner
function for flat space in Eq.\ (\ref{eq:Wig-delta}). In
(\ref{eq:newestestest-Wigner}), the integrand for $\pmbf\eta$ [recall
Eqs.\ (\ref{eq:x-y-etc}) and (\ref{eq:define-y})] will be significant
only when $\Re^D$ norms of the vectors fulfill
\bea
	|\xx\cosh\onehalf\tau\pm\yy\sinh\onehalf\tau|\ll R
			&\Rightarrow& \left\{\begin{array}{rcl}
		\abs\xx\cosh\onehalf\tau\ll R &\Rightarrow& \sinh\chi\ll1  \\
		\abs\yy\sinh\onehalf\tau\ll R &\Rightarrow& \sinh\tau\ll1
			\end{array}\right. \label{eq:implica1}\\
			&\Rightarrow&
				x\approx R(1,\chi\,{\pmbf\xi}),\quad
				y\approx R(\chi\,{\pmbf\xi}\cdot{\pmbf\eta},{\pmbf\eta}).
								\label{eq:implica2}
\eea
Also, using the limit in (\ref{eq:contraction-Phi}), and
approximating $\sinh\tau\approx\tau$ and
$\cosh\onehalf\tau\approx\cosh\chi\approx\cosh\omega\approx1$,
the Wigner function in Eq.\ (\ref{eq:newestestest-Wigner}) reduces to
\bea
	W_{\cal H}(f,g|\xx,\pp)&=& \frac{R^D}{(2\pi)^D}
	\int_0^\infty \tau^{D-1} \,d\tau
	\int_{{\cal S}^{D-1}} {d^{D-1}{\pmbf\eta}}
				\label{eq:newlimit-Wigner}\\ & &	{}\times
		f(x_0,\xx{-}\onehalf R\tau{\pmbf\eta})^*\,
		\exp(-iR\tau\,{\pmbf\eta}\cdot\pp)\,
		g(x_0,\xx{+}\onehalf R\tau{\pmbf\eta}).\nonumber
\eea
Finally, changing variables to $\zz=R\tau\,{\pmbf\eta}$ and noticing
that 
\be
	\int_{\Re^D} d^D\zz\,\cdots=R^D\,\int_0^\infty \tau^{D-1}\,d\tau\,
	\int_{{\cal S}^{D-1}}d^{D-1}{\pmbf\eta}\,\cdots
						\label{eq:integrals-dz}
\ee
we see that (\ref{eq:newlimit-Wigner}) reduces to
(\ref{eq:Wig-traditional}).

\subsection{Special case of one dimension}

	In the case $D=1$, the Wigner function
(\ref{eq:newestestest-Wigner}) actually coincides in form with the
corresponding one-dimensional standard flat space form
(\ref{eq:Wig-traditional}), as we now proceed to show.

	First, notice that the unit vectors \nn\ and $\pmbf\xi$ in Eq.\
(\ref{eq:Shapiro-functions}) are now the unit scalars $n,\xi=\pm1$,
and that the Shapiro functions become simple exponentials:
\be
	\Phi^\ssty{(1)}_p (R\xi\cosh\chi,R\xi\sinh\chi)
	= [\exp(-n\xi\chi)]^{-ipR}=\exp(in\xi pR\chi).
					\label{eq:one-dim-1}
\ee
We can now let $np\mapsto p$ and $\xi\chi\mapsto\chi$ with
$p,\chi\in(-\infty,\infty)$, and recognize that (\ref{eq:one-dim-1})
is a 1:1 function of only one variable, $x_1\in\Re$,
\be
	\Phi^\ssty{(1)}_p\of{x_1} = \exp(i\chi\,pR), \qquad x_1=R\sinh\chi.
					\label{eq:one-dim-2}
\ee
The arguments $x=(x_0,x_1)$ of the functions $f$ and $g$ in
(\ref{eq:newestestest-Wigner}) then simplify, in components, to
\be
	x=\vecdos{x_0}{x_1}\mapsto 
	x\cosh\onehalf\tau\pm y\sinh\onehalf\tau
		= R\vecdos{\cosh(\chi\pm\onehalf\eta\tau)}{
					\sinh(\chi\pm\onehalf\eta\tau)}.
							\label{eq:one-dim-3}
\ee
For short, we indicate $f(R\cosh\chi,R\sinh\chi)=f(\chi)$.
The unit vector $\pmbf\eta$ in
(\ref{eq:newestestest-Wigner})--(\ref{eq:define-y}) 
also becomes a unit scalar, $\eta=\pm1$, and the integral extends
over $y=\eta R(\sinh\chi,\cosh\chi)$, and $\Omega=\eta\chi$  [see
Eq.\ (\ref{eq:DDeltass})]. Finally, the integral over $\pmbf\eta$ reduces
to a sum over $\eta=\pm1$, and for
$\tau\mapsto\eta\tau\in(-\infty,\infty)$, the
Wigner function (\ref{eq:newestestest-Wigner}) becomes
\bea
	W_{\cal H}(f,g|x,p) 
	&=& \frac{R}{2\pi} \int_{-\infty}^\infty d\tau\,
		 f(\chi-\onehalf\tau)^*\, e^{-ipR\tau}\, g(\chi+\onehalf\tau)
							\label{eq:one-dim-4}\\
	&=& \frac{1}{2\pi} \int_{-\infty}^\infty d\upsilon\,
		 \widetilde f(p-\onehalf\upsilon)^*\, e^{+iR\upsilon x}\, 
			\widetilde g(p+\onehalf\upsilon).
							\label{eq:one-dim-4-p}
\eea
The last expression is the usual flat-space Wigner function
in terms of the conjugate wavefunctions on momentum space.
Finally, note that for $D=1$, the net multiplier which appears under
`boost' transformations in (\ref{eq:non-rel-covariance}) is unity, so
standard and Shapiro covariances coincide.


\section{Example: oscillator on the hyperbola}  \label{sec:4}

	We consider the open one-dimensional space which is
the upper branch of a hyperbola of fixed radius $R>0$,
\be
 {\cal H}^1_+=\{(x_0,x_1)\in\Re^2 \mid x_0^2-x_1^2=R^2\},
				\label{eq:hyperbola}
\ee
parametrized as usual by the hyperbolic angle $\chi\in\Re$.

\subsection{Laplace-Beltrami operator and the oscillator}

	When the potential $V\of\xx$ is a constant (corresponding
to a homogeneous optical medium), the $D=1$ Schr\"odinger equation
(\ref{eq:Schrodinger-osc-D}) is the free wave equation, and 
its Shapiro solutions are simply the oscillating exponentials
(\ref{eq:one-dim-2}), with energy $E=p^2/2\mu\ge0$.

	Since the Laplace-Beltrami operator on the hyperbola
(\ref{eq:hyperbola}) is $\Delta_\ssr{LB}=R^{-2} d^2/d\chi^2$, 
the Schr\"odinger equation for the conic oscillator
(\ref{eq:conicHO--potential}) is
\bea
	& &\hskip-70pt\Big(\frac{-1}{2\mu}\totder{^2}{\chi^2} 
				-R^2E_0\,\hbox{sech}^2\,\chi\Big) f(\chi)
		= R^2\,(E-E_0)\,f(\chi),\quad E_0=\onehalf\mu\omega^2R^2,
										\label{eq:Schrodinger}\\
	V(\chi)&=&\onehalf\mu\omega^2R^2 \frac{x_1^2}{x_0^2}
		=\sqrt{s(s+1)}\,(1-\hbox{sech}^2 \chi),
			\label{eq:P"osch-Teller-potential}\\
	  	s&=&-\onehalf+\sqrt{(\mu\omega R^2)^2+{\textstyle\frac14}}.
			\label{eq:def-of-s}
\eea
The bound solutions are \cite{Landau}:
\bea
	\psi_{n}^{s} (\chi) &=&\frac{2^{-(s-n)}}{\Gamma(s-n+1)}
			\,\sqrt{\frac{(s-n)\,\Gamma(2s-n+1)}{n!}}\nonumber\\
			& &\quad\times\hbox{sech}^{s-n}\chi\,\,
				_2F_1\left(\matrix{-n,\,\,2s-n+1\cr s-n+1}
					\Bigg|\frac{1-\tanh\chi}{2}\right)
								\label{eq:P-T-sol-hyp}\\
		&=&\sqrt{\frac{(s{-}n)\,n!}{\pi\,\Gamma(2s{-}n{+}1)}}
				\,\Gamma(s{-}n{+}\onehalf)\,
				(2\,\hbox{sech}\,\chi)^{s-n}\,
				C_n^{s-n+1/2}(\tanh\chi),\nonumber
\eea
where $n$ is a nonnegative integer bounded by $s+1$, and 
$C_n^{\alpha}(\xi)$ are the Gegenbauer (or ultraspherical) polynomials
\cite{Bateman} for $\alpha>-\onehalf$. The corresponding quantized
values of the energy are quadratic in $n$, and counted from the
lowest level up by 
\be
	E_{n}^s=\frac{\mu\omega^2 R^2}{2}-\frac1{2\mu R^2}(n-s)^2,\qquad
			n=0,1,2,\ldots\ < s+1.
							\label{eq:energies-POH}
\ee

	As a check on our concepts, we verify that the contraction limit
$R\to\infty$ of this system, when the radius of the hyperbola grows
without bound, is the harmonic oscillator on flat space.
Since the coefficient $s$ correspondingly grows as
$s\approx\mu\omega R^2$, the linear-quadratic spectrum of energies in Eq.\
(\ref{eq:energies-POH}) becomes the linear spectrum of the
quantum harmonic oscillator $E_n=\omega (n+\onehalf)$. To implement
this limit on the wavefunctions (\ref{eq:P-T-sol-hyp}), it is
convenient to use the following forms for the Gegenbauer polynomials
in $\xi=\tanh\chi$:
\be
	C_n^\alpha(\xi)=\left\{\begin{array}{rl}\displaystyle
	(-1)^{\frac12n}\frac{\Gamma(\alpha+\onehalf{n})}{
				(\onehalf{n})!\,\Gamma(\alpha)}
		\,\,_2F_1\left({-\onehalf{n},\ \onehalf{n}+\alpha\atop\onehalf}
				\Bigg|\ \xi^2\right),& n \hbox{ even},\\[5pt] \displaystyle
	(-1)^{\frac12(n{-}1)}\,\frac{\Gamma\Big(\alpha{+}\onehalf({n{+}1})\Big)}{
			\Big(\onehalf({n{-}1})\Big)!\,\Gamma(\alpha)} 
				\qquad\qquad\qquad\quad & \\
		\times 2\xi
		\,_2F_1\left({-\onehalf({n{-}1}),\ \onehalf(n{+}1){+}\alpha
											\atop\textstyle\frac32}
				\Bigg|\ \xi^2\right),& n \hbox{ odd}.
		\end{array}\right.  \label{eq:Gegenbauer-forms}
\ee
Then, for $\alpha=s-n+\onehalf\to\infty$, the hypergeometric polynomials
simplify: for $n$ even, $_2F_1(-\onehalf n,\alpha;\onehalf;\xi^2)
\approx {}_1F_1 (-\onehalf n;\onehalf;\alpha\xi^2)
=H_n(\sqrt\alpha \xi)$, and similarly for $n$ odd. Replacing this
into Eq.\ (\ref{eq:P-T-sol-hyp}), with
$\xi=\tanh\chi\approx\sinh\chi=x_1/R$ and
$\cosh^{-s+n}\chi\approx\exp(-s\tanh^2\onehalf\chi)$,
we obtain the harmonic oscillator wavefunctions on flat space,
\be
	\frac1{\sqrt R}\psi_{n}^{s}(\chi)
	\approx
	\frac1{\sqrt{2^n n!\, \sqrt{\pi/\mu\omega}}}\,e^{-{\mu\omega}x_1^2/2}
	\,H_n(\sqrt{\mu\omega}\,x_1).
						\label{eq:flatHo1}
\ee
The factor $\sqrt{R}$ restores the proper normalization on the $x_1$
axis. 

	In addition to the bound states, the sech-trough P\"oschl-Teller
potential also has free states with energy above the asymptotic value
of the potential $\lim_{\chi\to\pm\infty}V(\chi)=\onehalf\mu\omega^2R^2$.
These scattering solutions contain associated Legendre polynomials of
imaginary upper index:
\be
	\psi_p(\chi) = \frac{|\Gamma(1-ip)|}{2\pi}
			P_\sigma^{ip}(\tanh\chi), \qquad
		\begin{array}{l} p=R\sqrt{2\mu E-\mu^2\omega^2R^2}>0,\\
				\sigma=\onehalf\pm\sqrt{\omega^2\mu^2R^4
						+{\textstyle\frac14}}. \end{array}
									\label{eq:unbound}
\ee
These wavefunctions are Dirac-orthonormal.


\subsection{Momentum representation of the wavefunctions}

	The bound wavefunctions of the P\"oschl-Teller sech-trough
potential in the momentum representation, $\widetilde\psi_{n}^{s}(p)$
are found through the ordinary Fourier transform [Eqs.\
(\ref{eq:Shapiro-Mellin-tfm})--(\ref{eq:Shapiro-Mellin-inv}) for
$D=1$ and (\ref{eq:one-dim-2})] of the wavefunctions
$\psi_{n}^{s}(\chi)$ found in Eq.\ (\ref{eq:P-T-sol-hyp}). The result
is 

\vbox{\bea
	\widetilde\psi_{n}^{s}(p)&=& \sqrt{\frac{R}{2\pi}}\,
	\int_{-\infty}^{\infty}d\chi\,\exp(-ipR\chi)\,\psi_{n}^{s}(\chi) 
												\nonumber\\[2mm]
	&=&\frac{R}{2}
		\sqrt{\frac{\Gamma(2s-n+1)}{\pi(s-n)n!}}
		\frac{|\Gamma\Big(\onehalf(s-n-ipR)\Big)|^2}
                              {\Gamma(s-n)^2}\nonumber\\
	& &\quad\times{}_3F_2\left(
		{-n,\ 2s-n+1,\ \onehalf(s-n-ipR) \atop s-n+1,\ s-n}\Bigg|\ 1\right)
							\label{eq:Fou-tfmn-wavef-p}\\
		&=& \frac{(-i)^n\, R}{2\sqrt\pi}\,
			\frac{\sqrt{(s-n)\,n!\,\Gamma(2s-n+1)}
				   }{\Gamma(s)\,\Gamma(s+1)}\,
				   \left|\Gamma\Big(\onehalf(s-n-ipR)\Big)\right|^2
						\nonumber\\ & &\quad\times
			R_n\Big({-\onehalf}ipR;\onehalf(s{-}n),\onehalf(s{-}n),
						\onehalf(s{-}n),\onehalf(s{-}n){+}1\Big),
							\label{eq:Cont-Hahn-pols}
\eea}

\noindent where $R_n(z;\alpha,\beta,\gamma,\delta)$ 
are the continuous Hahn polynomials \cite{Hahn}.
On the other hand, the unbound solutions (\ref{eq:unbound}) are not
square-integrable, so their Fourier transform must be performed
allowing for the phase difference between asymptotic incoming and
outgoing waves, as determined by the scattering properties
of this P\"oschl-Teller potential. We shall not
further detail the free states here; they can be found in Ref.\
\cite{Basu-etal} among the coupling coefficients between the
$D^+\times D^-\to \sum D^+ +\int C$ irreducible representations
series of \SO{2,1}.

\subsection{Wigner function for the oscillator eigenstates
				on the hyperbola}

On the one-dimensional hyperbola ${\cal H}^1_+$, the Wigner function
(\ref{eq:Wig-new}) collapses to
(\ref{eq:one-dim-4})--(\ref{eq:one-dim-4-p}), its usual form in 
Eqs.\ (\ref{eq:Wig-traditional})--(\ref{eq:Wig-delta}) for $D=1$. 

	We are interested in the single-function form $W(f|\xx,\pp)\equiv
W(f,f|\xx,\pp)=W(\widetilde f,\widetilde f|\pp,-\xx)$ for the
P\"oschl-Teller wavefunctions, whose explicit form is in Eq.\
(\ref{eq:P-T-sol-hyp}) for $\psi_{n}^{s}(x)$, and in Eq.\
(\ref{eq:Fou-tfmn-wavef-p}) for $\widetilde\psi_{n}^{s}(p)$; we find
the latter more amenable to analytic solution. We change the
integration to a contour along the imaginary axis, and find

\vbox{\bea
	W(\psi_{n}^{s} |\chi,p) \!\!
	&=& \!\!\frac{R^2}{8\pi^2} \frac{s{-}n}{n!\,\Gamma(2s{-}n{+}1)}
	\sum_{m,l=0}^{n}
	\frac{(-n)_m\,(-n)_l}{\Gamma(s{-}n{+}m)\,\Gamma(s{-}n{+}l)}
										\nonumber\\[2mm]
	& &{\!\!}\times \frac{\Gamma(2s{-}n{+}m{+}1)\,\Gamma(2s{-}n{+}l{+}1)}{
		\Gamma(s{-}n{+}m{+}1)\,\Gamma(s{-}n{+}l{+}1)}\,\frac{1}{l!\,m!}
			\,I^s_n (\chi, p), \label{eq:Woftilde-Poschl-Teller1}\\
	\hbox{where \ }I^s_n (\chi, p)\!\!
	&=&{}\!\!{-} \frac{4i}{R}\int^{i\infty}_{-i\infty}dz\,
		\Gamma(\onehalf\of{s{-}n}{-}\onehalf{ipR}{-}z)\,
		\Gamma(\onehalf\of{s{-}n}{+}\onehalf{ipR}{+}z{+}m) \nonumber\\[2mm]
			& &{\!\!}\times e^{-4\chi z}\,
				\Gamma(\onehalf\of{s{-}n}{+}\onehalf{ipR}{-}z)\,
				\Gamma(\onehalf\of{s{-}n}{-}\onehalf{ipR}{+}z{+}l).
							\label{eq:Woftilde-Poschl-Teller2}
\eea}

\noindent The last integral can be computed in the complex plane
straightforwardly, and leads to a pair of complex conjugate
$_2F_1$-functions of $e^{-4\chi}$. We thus finally have

\vbox{\bea
	W(\psi_{n}^{s}|\chi,p) 
	&=&\frac{2R\,(s{-}n)\,n!\,e^{-2\chi(s-n)}}{
			\pi\Gamma(2s{-}n{+}1)} \nonumber\\[2mm]
	&\times &	\sum_{m,l=0}^{n}  
		\frac{\Gamma(2s-n+m+1)\,\Gamma(2s-n+l+1)}{
			\Gamma(s{-}n{+}l+1)\,\Gamma(s{-}n{+}m{+}1)\,\Gamma(s-n+l)}
										\nonumber\\[2mm]
	&\times & \frac{(-1)^{l+m}}{l!\, m!\,(n-m)!\,(n-l)!} 
						\label{eq:Woftilde-Poschl-Teller3}\\[2mm]
	&\times &\hbox{Re}\,\Bigg\{\Gamma(ipR)\,\Gamma(s{-}n{+}l{-}ipR)\,
						e^{2ipR\chi} 	\nonumber\\[2mm]
	&\times &			{}_2F_1\left({s-n+m,\ s-n+l-ipR,
			\atop 1-ipR}\Bigg|\ e^{-4\chi} \right)\Bigg\}.\nonumber
\eea}

\noindent We note that this expression is suitable for numerical
computation only for $\chi>0$ because it converges fast, but it holds
everywhere analytically, with the reflection symmetries 
$W(\psi_{n}^{s}|\chi,p)=W(\psi_{n}^{s}|{-\chi},p)
=W(\psi_{n}^{s}|\chi,{-p})$. 

  The Wigner functions, together with their
marginal projections $|\psi_{n}^s(x)|^2$ and $|\widetilde
\psi_{n}^s(p)|^2$, are shown in Figures 1, for $n=0,1,2,3$, and
for the potential depth parameter $s=4$ and 30. It can be appreciated
that the Wigner function of the most tightly bound states resemble
the familiar Gaussian-bell form of the harmonic oscillator ground
state. According to (\ref{eq:energies-POH}), for $s=4$ there are only
5 bound states ($n=0,\ldots,4$), and as the energy of the state
approaches the binding energy, the wavefunction stretch\-es in space
with ever-smaller momentum in a neighborhood of the classical turning
point. The contraction limit can be appreciated in the $s=30$ column,
corresponding to a large binding energy; the Wigner functions for the
eigenstates in Eq.\ (\ref{eq:P-T-sol-hyp}) approach the familiar
Laguerre-Gaussian form that corresponds to the Wigner function of the
harmonic oscillator wavefunctions on flat space.


\section{Concluding remarks}  \label{sec:5}

	We have generalized the Wigner quasiprobability distribution
function by replacing the oscillating exponential functions of the
standard version, which are Dirac solutions of the free Schr\"odinger
equation on flat space, by the Shapiro functions, because they are
solutions to the Laplace-Beltrami equation on a simply connected
hyperbolic space, while respecting the midpoint condition through an
appropriate Dirac-like $\Delta^D(x;x',x'')$. The role of the Fourier
transform in the standard version is transfered by the Shapiro
functions to a Mellin-like transform between the position and
momentum coordinates of phase space. Indeed, the relation between
what were called position and momentum variables is actually defined
by the argument and index of the basis of Shapiro functions. Thus
built, the proposed Wigner function is covariant under the group
$\SO{D,1}$ of motions of the hyperbola, with the hyperbolic
translations extracting a multiplier factor. The correct marginals
are found and the contraction limit to flat space returns the
standard Wigner function.

	The transformations of the momentum direction under translations
of the hyperbola are the (unique) action of $\SO{D,1}$ on the sphere
${\cal S}^D$ \cite{Bargmann}. There are several models of Hamiltonian
systems where momentum is restricted to a sphere, such as geometric
and Helmholtz (monochromatic) optics \cite{KBW-GK,KBW-MAA-GWF}. In
the first model, a `Descartes sphere' of momentum vectors \pp\ which
is of radius $\abs\pp=n\of\xx$ (the refractive index), is associated
to each point \xx\ in space.  This sphere of momentum vectors has
been subject to Bargmann's `boost' transformation in Refs.\
\cite{Rel-Coma} (which is unique for Lorentz groups acting on
spheres) and here given by (\ref{eq:nnprime-perp-parallel}), with a
canonically conjugate transformation of the ray positions at a plane
screen. The resulting phenomenon has been called relativistic coma,
because the aberration in the images projected on moving screens in
vacuum is comatic. The Doppler effect is of course absent, so this
transformation cannot be called physically relativistic in the
Helmholtz case, which involves a configuration space with a nonlocal
metric. Indeed, the difference between the various models consists in
their space of positions; here, it is the hyperboloid.  A Wigner
function on spheres will be examined in part II of this title; it is
expected to clarify further the use of function bases to define
conjugate variables as a simile or substitute for phase space. 

	 We point out that the example of the one-dimensional Wigner
function of P\"oschl-Teller wavefunctions is of interest not only for
the traditional quantum mechanical model, but also for the paraxial
propagation of light along shallow non-harmonic waveguides whose
index of refraction has a ${\rm sech}^2$ profile, as given by Eq.\
(\ref{eq:nyV}).  Finally, one of the manifestations of higher
symmetry is the appearance of `closed-form' wavefunctions expressed
in terms of well-known (and some not-so-well known) special
functions. Generally, the Fourier transforms and Wigner functions of
such wavefunctions are again known special functions because
symmetry, when it occurs, is displayed in phase space.  


\section*{Acknowledgements} We are indebted with Natig M.\
Atakishiyev for his interest and comments on our work, and to Rufat
Mir-Kasimov for a crucial remark on the Shapiro basis.  We thank the
support of the Direcci\'on General de Asuntos del Personal
Aca\-d\'e\-mico, Universidad Nacional Aut\'onoma de M\'exico ({\sc
dgapa--unam}) grant IN112300 {\it Optica Matem\'atica}.
G.S.P.\ acknowledges the Consejo Nacional de Ciencia y
Tecnolog\'{\i}a (M\'exico) for a C\'atedra Patrimonial Nivel II, and
the Armenian National Science and Engineering Foundation for grant
PS124--01. 



\newpage

\noindent {\LARGE  Figure caption}

\vfil

\bigskip\bigskip\hrule\bigskip

\noindent{\bf Figure 1.}\quad Wigner functions of the P\"oschl-Teller
eigenfunctions $\psi_n^s(\chi)$ on a quadrant of phase space (axes
are position $\chi\sqrt{s}$ and momentum $pR/\sqrt{s}$; the quadrants
have reflection symmetry across the axes). Rows are numbered by the
mode $n=0,1,2,3$. Left: $s=4$ (so only five states, $n=0,\ldots,4$,
are bound); right: $s=30$ (so states are bound up to $n=30$).  White
is the maximum, black is the minimum; the shade at the upper right
corner corresponds to zero.  From each Wigner function we
project up the marginal distribution of position
$|\psi_n^s(\chi)|^2$, and right the marginal of momentum
$|\widetilde\psi_n^s(p)|^2$.

\bigskip\hrule\bigskip\bigskip

\vfil

\null\eject
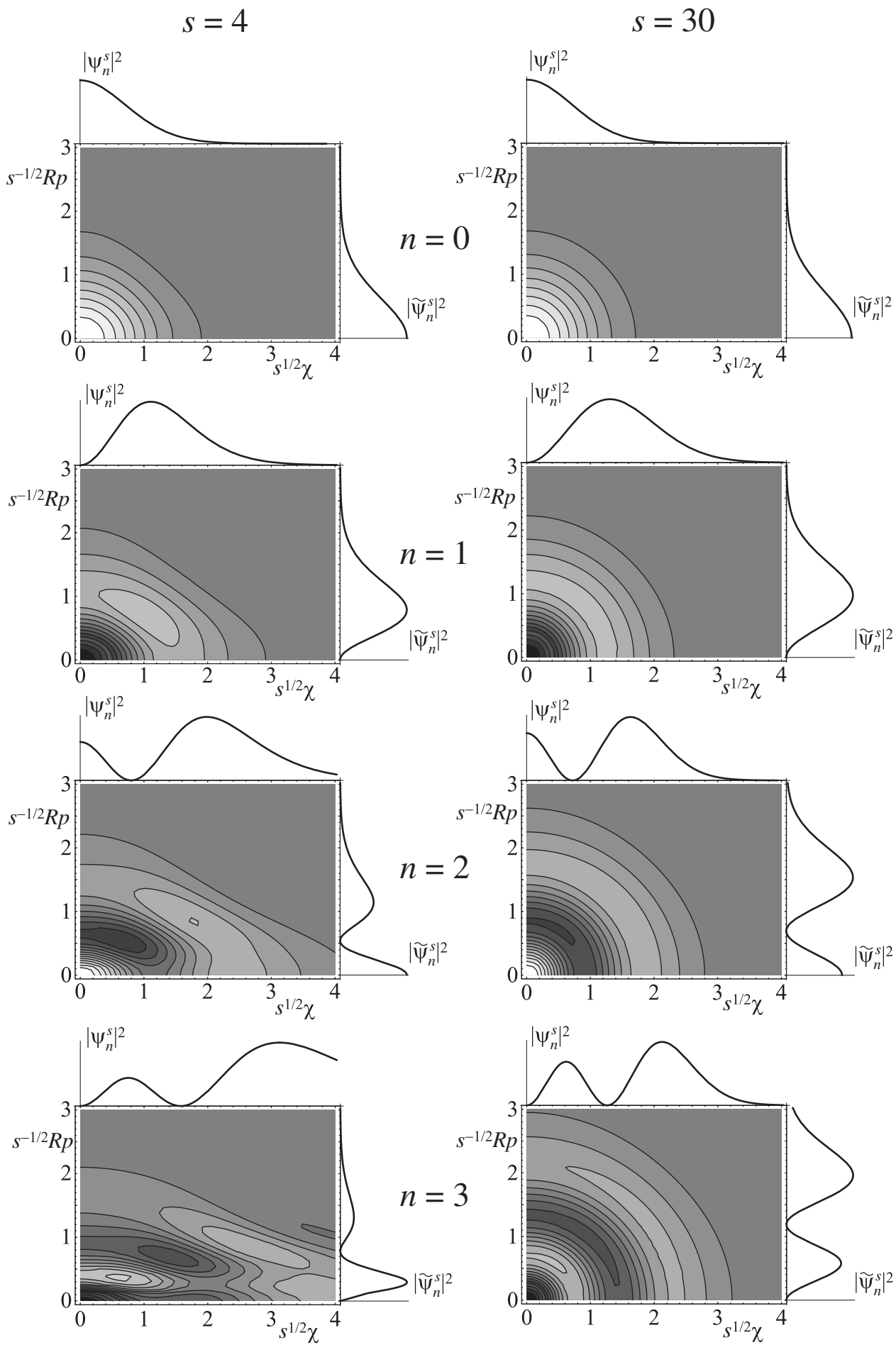
\end{document}